# A Monte Carlo study of O(3) antiferromagnetic models in three dimensions

J. L. Alonso, A. Tarancón,
*Departamento de Física Teórica, Facultad de Ciencias, Universidad de Zaragoza, 50009 Zaragoza, Spain*
(e-mail buj@cc.unizar.es, tarancon@sol.unizar.es)

H. G. Ballesteros, L. A. Fernández,
V. Martín-Mayor and A. Muñoz Sudupe
*Departamento de Física Teórica I, Facultad de Ciencias Físicas, Universidad Complutense de Madrid, 28040 Madrid, Spain*
(e-mail hector, laf, victor, sudupe@lattice.fis.ucm)

July 17, 1995

**Abstract**

We study three antiferromagnetic formulations of the O(3) spin model in three dimensions by means of Monte Carlo simulations: 1. a two parameter $\sigma$ model with nearest and next to nearest neighbors couplings in a cubic lattice; 2. a face centered cubic lattice with nearest neighbors interaction; 3. a cubic lattice with a set of fully frustrating couplings. We discuss in all cases the vacua properties and analyze the phase transitions. Using Finite Size Scaling analysis we conclude that all phase transitions found are of first order.

# 1 Introduction

There are several motivations to study antiferromagnetic O(3) $\sigma$ models in three dimensions.

The first one is in the framework of classical spin models that are of interest for systems of high spin particles. Most of the materials with only one kind of magnetic ions have an ordered antiferromagnetic phase at low temperature. These materials are generally ionic crystals (oxides, chlorides, fluorides, ... ) in which the magnetic ions are surrounded by anions, its interactions being of superexchange type [1].

The second one comes from High Energy Physics. In four dimensions the Quantum Field Theory is nonperturbatively well established only in the case of asymptotically free fields. For some models such as $\lambda\phi^4$ it has been shown that they are trivial (non interacting) when the nonperturbative effects are properly taken into account. Although the results for more complex systems (scalars or fermions coupled to gauge fields) are not conclusive, the *Triviality Problem* is without doubt one of the main open questions in the subject.

It is unclear the role that the antiferromagnetic models could play in the formulation of Relativistic Quantum Field Theories. However, the antiferromagnetic models have a very rich phase space and presumably could present new universality classes with alternative formulations of continuum quantum field theories [2, 3].

The study of simple versions such as three dimensional $\sigma$ models may be useful as a step towards the four dimensional theory.

Finally, and perhaps the strongest motivation is their close connection with models which describe high $T_c$ superconductors [4, 5].

The low temperature behavior of a 2 dimensional quantum Heisenberg antiferromagnet (2-DQHA) can be described by a nonlinear $\sigma$ model in (2+1) dimensions [4]. This result may be obtained from the Weinberg Theorem [6]: the low energy physics of a theory with a global symmetry group G spontaneously broken to a subgroup H coincide at leading order in momenta with that of the nonlinear $\sigma$ model defined on the coset space G/H. In our case, G=SU(2) (spin 1/2) and H=U(1) (rotations around the direction chosen by the system). Hence, the group manifold is $S^2 \sim SU(2)/U(1)$, yielding the O(3) nonlinear $\sigma$ model in the low-energy, long-wavelength limit. The application of the Weinberg theorem to antiferromagnets is due to Johannesson [7].

Applying this approach to the spin-$\frac{1}{2}$ Heisenberg model on a square lattice, with only nearest neighbors interactions, Chakravarty, Halperin and Nelson [4] obtained a result for the correlation length which is in good agreement with the $La_2CuO_4$ data, the parent compound of the first high-$T_c$ superconductor discovered [8]. Subsequent experiments on other compounds have shown that two-dimensional antiferromagnetic correlations are indeed a hallmark of these copper-oxide phase (before they become superconductors).

To make the copper-oxides superconducting one must *dope* the parent compound. In the case of $La_2CuO_4$, this means that one replaces some fraction of La with Ba, Ca or Sr. As an effect, the unpaired valence electrons in the hybridized 3d-3p band go somewhere else, leaving a *hole* behind. These holes lead to the destruction of long range order. The resulting state is still not fully understood.

A frustrating interaction can be added to the 2-DQHA playing a role similar to that of doping the system (within an adiabatic approximation [9]). In both cases the Néel ordered state gets perturbed, and one observes a disordering transition when the perturbation becomes sufficiently large [7, 10]. The simplest frustrated Heisenberg Antiferromagnet in (2+1) dimensions is the $J_1 - J_2$ model defined on a square lattice by

$$H = J_1 \sum_{\mathrm{nn}} \boldsymbol{S}_i \cdot \boldsymbol{S}_j + J_2 \sum_{\mathrm{nnn}} \boldsymbol{S}_i \cdot \boldsymbol{S}_j \ , \tag{1}$$

where we denote by "nn" the nearest neighbors and by "nnn" the second nearest neighbors (or next to nearest neighbors). In this Hamiltonian, the $J_1$ term describes the usual Heisenberg interaction of nearest neighbors spins ($S = 1/2$) on the square lattice, while the $J_2 < 0$ one introduces a frustrating interaction between next nearest neighbors sites.

With frustration replacing doping, charge is thrown away and clearly one can not explain superconductivity in this way. The hope is rather to find some clues or insights about the physics of two-dimensional disordered antiferromagnets which can be carried back to the *real problem*. Independently of this, the zero-temperature collapse of the Néel state is an archetype of quantum phase transition and is well worth a study on its own merits.

The frustrated magnetic systems are interesting in themselves, in the light of numerous theoretical predictions on the nature of the disordered ground state in quantum spin systems [11, 12]. For instance, antiferromagnets on a squared lattice, which are frustrated by adding second and third neighbors couplings [13], show interesting phases with incommensurate, planar and spiral correlations.

Also weakly frustrated $S = 1/2$ Heisenberg antiferromagnets in two dimensions can be mapped onto a nonlinear $\sigma$ model in the continuum limit (*i.e.*, for large wavelengths) [7]. So, it looks natural to study the phase diagram and the nature of the disordered phase of the frustrated nonlinear $\sigma$ model, although, until now, only the non-frustrated nonlinear $\sigma$ model has been used in the mapping.

# 2 Definition of the models

The simplest O(3) $\sigma$ model in a cubic lattice is described by the action

$$-S = \beta \sum_{\mathrm{nn}} \boldsymbol{\sigma}_i \cdot \boldsymbol{\sigma}_j \ , \tag{2}$$

where $\boldsymbol{\sigma}_i$ is a three components normalized real vector. For $\beta < 0$, the antiferromagnetic system is trivially related by a ferromagnetic one with the transformation

$$\{\boldsymbol{\sigma}(x,y,z)\};\ \beta \longrightarrow \{\boldsymbol{\sigma}(x,y,z)(-1)^{x+y+z}\};\ -\beta \ . \tag{3}$$

Thus, we have to go beyond the naive definition of an antiferromagnetic $\sigma$ model in a simple cubic lattice.

There are several procedures to break the symmetry (3) or to make the vacuum frustrated. Without the addition of new fields we can:

1. Add new couplings to the usual nearest neighbors coupling.

2. Change the geometry of the lattice.

3. Use different couplings along the lattice.

We will discuss these three possibilities in the following subsections.

## 2.1 The two parameter model

The simplest multiparameter model corresponds to the addition of a next nearest neighbors coupling. Consider then the two parameter action

$$-S = \beta_1 \sum_{\mathrm{nn}} \boldsymbol{\sigma}_i \cdot \boldsymbol{\sigma}_j + \frac{1}{2}\beta_2 \sum_{\mathrm{nnn}} \boldsymbol{\sigma}_i \cdot \boldsymbol{\sigma}_j \ , \tag{4}$$

where the first sum extends over nearest neighbors pairs while the second corresponds to next to nearest neighbors ones (distance $\sqrt{2}a$). The factor $\frac{1}{2}$ in the second term is added for convenience.

Notice that under the staggered transformation (3) the second sum in (4) does not change so it is not possible to map the negative $\beta_2$ values onto positive ones.

As a general rule, we shall refer to the adimensional quantities $\beta_i/kT$ as $\beta_i$, as they are the only relevant quantities for simulation.

Of course, new couplings may be added indefinitely. In *ferromagnetic* models, the new couplings can be useful in the framework of Renormalization Group Studies to move the simulation point closer to the fixed point; variations in the critical behavior of these systems are not expected with these additions (*Universality*). It is an open question how the addition of further couplings can affect the first order transitions that we found in *antiferromagnetic* models. This is beyond the scope of this paper, but let us remark that we consider very interesting a deep study of this subject specially using Renormalization Group Techniques.

## 2.2 The FCC lattice

Most of the work in Monte Carlo simulations is done in simple cubic lattices, but it is possible to choose lattices with translational and rotational symmetries that avoid the staggered degeneration. In 3 dimensions there are several cubic lattices that preserve the symmetry under $\pi/2$ rotations: the face centered cubic (FCC), the single site interior centered cubic (BCC) , and the tetrahedrical (diamond). Of them, only the FCC breaks the staggered degeneration for nearest neighbors interactions: it is easy to realize that a body centered lattice (BCC) can be seen as two interpenetrating single cubic lattices. This means that, by changing the spin signs in one of the sublattices, every antiferromagnetic bond becomes a ferromagnetic one, so that we expect it to belong to the same universality class as the usual nonlinear $\sigma$ model. An analogous argument can be set for the tetrahedrical lattice. Of course one could change this by adding further couplings.

When we choose $\beta_1 = 0$ in (4), defined on a cubic lattice, two sublattices are decoupled. Each one of them is FCC, which is not bipartite, that is, it can not be further separated in two non interacting sublattices.

### 2.3 The Fully Frustrated model

By choosing different couplings for each site, the vacuum may become frustrated. Models with random couplings present a spin glass behavior. In this paper we limit ourselves to models with a regular action. So, as a third way of breaking the symmetry in (3), we study the Fully Frustrated (FF) model that corresponds to a selection of the sign of the coupling in a regular way but producing frustrated vacua.

The interaction in this model is defined only between nearest neighbors, but the coupling sign alternates so that every plaquette in the cubic lattice has an odd number of antiferromagnetic couplings. Therefore the ground state is frustrated.

We define $\beta_{x,y,z;\mu}$ as the coupling of the link pointing in the $\mu$ direction from the $x, y, z$ lattice site. A possible definition of a Fully Frustrating set of couplings is

$$\begin{aligned} \beta_{x,y,z;0} &= \beta(-1)^{x+y} \ , \\ \beta_{x,y,z;1} &= \beta(-1)^{z} \ , \\ \beta_{x,y,z;2} &= \beta \ , \end{aligned} \tag{5}$$

where the values $\mu = 0, 1, 2$ correspond to the $x, y, z$ directions respectively. Notice that there is a symmetry of the action when changing the sign of the spin at a site and simultaneously changing the sign of the couplings at the links starting in that site ($Z_2$ local gauge symmetry). This property allows a large flexibility when selecting the sign of the couplings.

# 3 The method

### 3.1 Monte Carlo algorithms

For the updating we have used mainly a Metropolis algorithm followed by several, typically nine, overrelaxation steps. After thermalization, the number of Monte Carlo sweeps performed for each simulation has been of the order of $10^6$ in the largest lattices. In all cases the autocorrelation time is much smaller than the total Monte Carlo time used for measures.

We represent the O(3) variables as three real numbers. It is more efficient, to speed up the computation, to evaluate directly the third component of the vectors rather than using the normalization constraint.

The overrelaxed microcanonical update [14] has a very simple implementation in these models, since it only requires sums and products of real numbers. Let us call $\boldsymbol{v}$ the terms in the action which multiply the variable $\boldsymbol{u}$ that we want to update:

$$\boldsymbol{v} = \beta_1 \sum_{\rm nn} \boldsymbol{\sigma}_i + \frac{1}{2}\beta_2 \sum_{\rm nnn} \boldsymbol{\sigma}_i \ . \tag{6}$$

Then, the transformation we use corresponds to the maximum spin change without modifying the energy, that is:

$$\boldsymbol{u} \longrightarrow 2\frac{\boldsymbol{v}\cdot\boldsymbol{u}}{\boldsymbol{v}\cdot\boldsymbol{v}}\boldsymbol{v} - \boldsymbol{u} \ . \tag{7}$$

The update is made sequentially on the $x$, $y$ and $z$ axes for largest efficiency (the dynamic exponent becomes 1).

In the Metropolis algorithm, to compute the tentative change, we calculate an uniformly distributed vector inside a sphere of radius $\delta$ and we add it to the original spin variable, normalizing afterwards. We perform 3 hits per update, selecting the $\delta$ value to ensure an acceptance rate over a 75%

We have also implemented a Wolff's single cluster algorithm [15] but, due to the frustration, the cluster size represents a very large fraction of the total lattice volume when we are near to the antiferromagnetic transition, making the algorithm very inefficient. For this reason we have used it only to study the ferromagnetic transition.

### 3.2 Observables

We measure the energies

$$E_1 = \frac{1}{3V}\sum_{\rm nn} \boldsymbol{\sigma}_i \cdot \boldsymbol{\sigma}_j \ , \tag{8}$$

$$E_2 = \frac{1}{6V}\sum_{\rm nnn} \boldsymbol{\sigma}_i \cdot \boldsymbol{\sigma}_j \ , \tag{9}$$

where $V$ is the lattice volume. With these definitions we ensure that $E_1, E_2 \in [-1, 1]$. For the FCC lattice we measure only the first neighbor energy. In the FF case we must consider the sign of the coupling if $i$-$j$ corresponds to a signed link.

The standard magnetization is defined as:

$$\boldsymbol{M} = \frac{1}{V}\sum_i \boldsymbol{\sigma}_i \ . \tag{10}$$

In the case of antiferromagnetic phases the previous quantity vanishes so it is not an order parameter. In the two parameter model as well as in the

FCC lattice we can define an order parameter as the set of vectors (staggered magnetization)

$$
\begin{aligned}
\boldsymbol{M}_x^s &= \frac{1}{V}\sum_i (-1)^x \boldsymbol{\sigma}_i \ , \\
\boldsymbol{M}_y^s &= \frac{1}{V}\sum_i (-1)^y \boldsymbol{\sigma}_i \ , \\
\boldsymbol{M}_z^s &= \frac{1}{V}\sum_i (-1)^z \boldsymbol{\sigma}_i \ ,
\end{aligned}
\tag{11}
$$

where $x, y, z$ are the coordinates of $i^{th}$ site of the lattice. Notice that, for instance, $\boldsymbol{M}_x^s$ is just the Fourier transform of the spatial spin distribution at momentum $(\pi, 0, 0)$.

For the fully frustrated model, we have computed the following set of vectors:

$$
\boldsymbol{M}_p(i,j,k) = \frac{8}{V} \sum_{\substack{x,y,z \\ \text{(even)}}} \boldsymbol{\sigma}(x+i, y+j, z+k), \quad i,j,k = 0,1 \ . \tag{12}
$$

Using the local gauge invariance it is easy to check that the mean values of $\boldsymbol{M}_p(i,j,k)$ do not depend of $i,j,k$. We shall refer to this common expectation value as the *period two magnetization.*

In finite lattices all the vector magnetizations have zero mean values. In practice we measure the magnetization squared from which we can obtain $\langle \boldsymbol{M}^2 \rangle$, $\langle \boldsymbol{M}^4 \rangle$, $\langle |\boldsymbol{M}| \rangle$, etc., where $\boldsymbol{M}$ is one of the magnetizations defined above.

From the the mean values of functions of the magnetizations squared we compute the Binder cumulant and the susceptibility.

The Binder cumulant is defined as

$$
U_L = 1 - \frac{\langle \boldsymbol{M}^4 \rangle}{3 \langle \boldsymbol{M}^2 \rangle^2} \ , \tag{13}
$$

and it is of interest to determine the transition point as well as some quotients of critical exponents in second order phase transitions.

Regarding the susceptibility. We use the definition

$$
\chi = V \left( \langle \boldsymbol{M}^2 \rangle - \langle |\boldsymbol{M}| \rangle^2 \right) \ . \tag{14}
$$

Another very interesting quantity is the correlation length. The usual definition looking at the exponential tail of the propagators at large separation is very difficult to measure due to fluctuations. Also we would need asymmetric lattices with one dimension longer than the others to be able to observe the exponential tail.

We have used instead the second momentum definition considered in ref. [16]. From the propagator

$$
G(\boldsymbol{r}_i - \boldsymbol{r}_j) = \langle \boldsymbol{\sigma}_i \cdot \boldsymbol{\sigma}_j \rangle \ , \tag{15}
$$

we compute the Fourier transform at zero momentum ($g_0$) and also at minimal non zero momentum $|\boldsymbol{p}_{\min}| = 2\pi/L$ (we call it $g_1$), $L$ being the lattice length. We use the definition for the correlation length

$$\xi = \left(\frac{g_0/g_1 - 1}{4\sin^2(\pi/L)}\right)^{1/2} . \tag{16}$$

We address to reference [16] for details. For the antiferromagnetic vacua the previous definition is appropriately generalized using the staggered magnetization.

The definition (16) has the same Finite Size Scaling behavior that the exponential correlation length, but it is easier to measure. When there is spontaneous magnetization (16) makes no sense as a correlation length (it always grows with $L$).

In practice we store the values of the energies and magnetizations of configurations usually separated by 10 Monte Carlo sweeps. From this data we can compute the derivatives with respect to the couplings as the connected correlation with the energies. We define the specific heat matrix as

$$C_{i,j} = \frac{\partial E_i}{\partial \beta_j} = 3V\left(\langle E_i E_j\rangle - \langle E_i\rangle\langle E_j\rangle\right) \tag{17}$$

The derivatives of functions of the magnetization can be computed in the same way. In addition, we use the spectral density method [17] for calculating the observables in a neighborhood of the transition point.

# 4 Finite Size Scaling

To study the properties of the transition we have used a Finite Size Scaling Analysis.

The correlation length exponent $\nu$ can be measured from quantities which have a $L^{1/\nu}$ scaling behavior, like the maxima of $d\log\langle|\boldsymbol{M}|\rangle/d\beta$ or $(dU_L/d\beta)$. The former quantities will be used in this work. Alternatively, we can use the same quantities evaluated at the infinite volume critical temperature, if it is known at all. We can estimate this value in second order phase transitions by measuring the Binder cumulant for various lattices and locating the point where the graphics cross. The scaling behavior of the crossing point, $\beta_{L_1,L_2}$ for lattices $L_1$, $L_2$ was obtained by Binder [18]:

$$\frac{1}{\beta_c} - \frac{1}{\beta_{L_1,L_2}} \sim \frac{1}{\log(L_1/L_2)} . \tag{18}$$

The exponents $\alpha$, and $\gamma$ can be obtained computing the maximum of the specific heat and susceptibility respectively:

$$C \sim L^{\alpha/\nu}, \tag{19}$$

$$\chi \sim L^{\gamma/\nu}, \tag{20}$$

For computing the $\beta$ exponent we measure the value of the magnetization at the transition point using the relation

$$\langle |\boldsymbol{M}| \rangle_{\beta_c} \sim L^{-\beta/\nu} \ . \tag{21}$$

However at a first order phase transition the correlation length does not diverge. Nevertheless, there is a scaling behavior with fictitious critical exponents.

For example, the specific heat grows with the volume $L^d$ of the system if the lattice size $L$ is large enough, we will say then that $\alpha/\nu = d$. Summarizing, the *critical exponents* of a first order phase transition obtained from Finite Size Scaling are[19]

$$\nu = \frac{1}{d} \ , \quad \alpha = 1 \ , \quad \gamma = 1 \ . \tag{22}$$

Most of these scaling relationships are for the maximum of a given observable. But different observables do not necessarily have their maxima at the same value of the coupling. This means that short trial runs are needed to locate the apparent critical coupling for one of them (for instance the specific heat or the susceptibility) and, then, rely on the standard spectral density method [17] to extrapolate to other values of the couplings and to obtain the maxima of different observables or their values at a suitable point.

# 5 Numerical Results

## 5.1 The two parameter model

We have studied the parameter region $\beta_1 > 0$. The case $\beta_1 < 0$ can be mapped onto the $\beta_1 > 0$ one with the transformation (3). The $\beta_1 = 0$ limit is a special case as two sublattices are decoupled, so we will consider it in the following subsection as an independent model (FCC lattice).

We have found three different phases: the ferromagnetic phase, that covers most of the region of positive coupling values, the antiferromagnetic phase, for large enough negative values of $\beta_2$, and an intermediate disordered (paramagnetic) region.

The order parameter for the paramagnetic-ferromagnetic transition is the usual magnetization (10). For the paramagnetic-antiferromagnetic transition we use the staggered magnetization (11) as an order parameter, since the interaction stabilizes a structure of alternate planes (see below).

The corresponding transition lines are depicted in figure 1. These lines have been obtained from simulations in small lattices ($L \leq 16$). The finite size effects in the apparent critical points are negligible at the scale of figure 1. The Finite Size Scaling analysis reported below has been done at the points labeled **A** to **F**.

We have analyzed three points (**A**, **B** and **C** in figure 1) along the transition line separating the disordered from the ferromagnetic phase. One of them (**B**) is the standard $\sigma$ model critical point, **A** is the FCC ferromagnetic lattice (at $\beta = 0.619(5)$), and the last one (**C**) is at $\beta_1 = 2.0$ and $\beta_2 = 0.853(5)$.

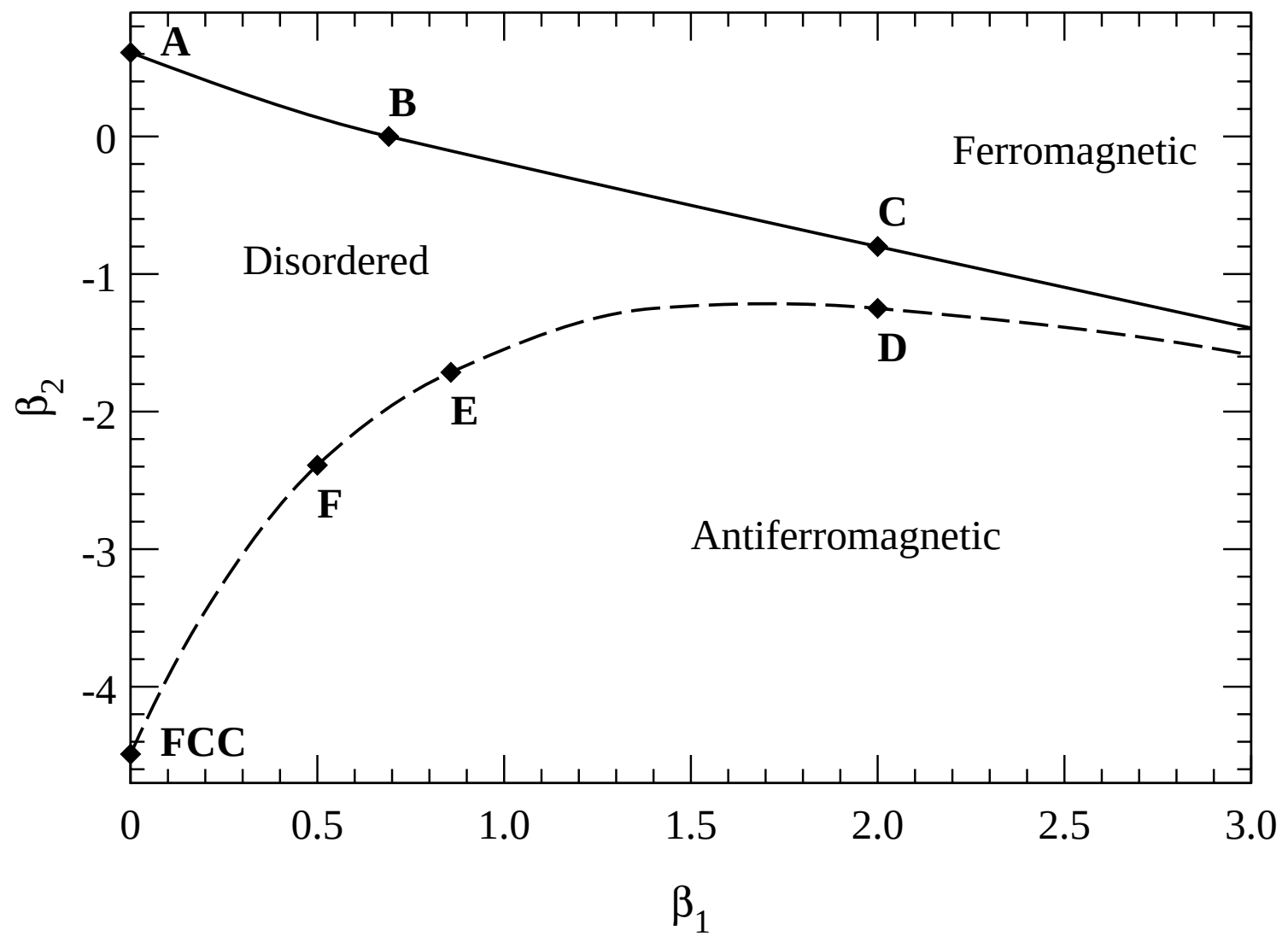

Figure 1: Phase diagram of the two parameter model. We plot black diamonds at the points referred on the text. The lines correspond to the ferromagnetic (solid) and antiferromagnetic (dashed) transitions.

At every point, we have measured the critical exponents and we have checked that they agree well with known values for the standard $\sigma$ model [20]; we remark that the exponent $\nu$ has been measured within a 2% of accuracy. We can say with a high level of confidence that all the paramagnetic-ferromagnetic transition lies in the same universality class.

The second line separates a disordered paramagnetic phase from an antiferromagnetic ordered phase. The ground state of this phase is given by one of these three types of configurations, ($\boldsymbol{u}$ being an arbitrary unit length vector):

$$\begin{aligned} \boldsymbol{\sigma}(x,y,z) &= (-1)^x \boldsymbol{u} \ , \\ \boldsymbol{\sigma}(x,y,z) &= (-1)^y \boldsymbol{u} \ , \\ \boldsymbol{\sigma}(x,y,z) &= (-1)^z \boldsymbol{u} \ . \end{aligned} \tag{23}$$

Consequently the O(3) symmetry is spontaneously broken and the magnetization $\boldsymbol{M}^s$ defined in (11) is adapted to this order. Notice also that the spatial rotational (cubic group) invariance is also broken since the vacua are anisotropic.

We have studied in detail three points along the antiferromagnetic line (labeled **D**, **E** and **F** in figure 1). We have measured the specific heat, the susceptibility, the magnetization $\boldsymbol{M}^s$ and the correlation length in each one. We will report the results in the following paragraphs.

The specific heat is defined as the fluctuation of the energy or equivalently the derivative of the energy with respect to the coupling. As we work with two couplings ($\beta_1$, $\beta_2$), and their conjugate energies ($E_1$ and $E_2$ respectively), the specific heat is really a $2 \times 2$ matrix. For simplicity we just present the results for the derivative of $E_2$ respect to $\beta_2$ in the case of point **D**; and for points **E** and **F** we measure the derivative of $(E_1 - 2E_2)$ with respect to the linear combination $\beta_1 - \frac{1}{2}\beta_2$. We have checked that derivatives in other directions scarcely add any new information.

Using the Spectral Density Method we measure the specific heat at the point where it reaches its maximum value along some straight line in the $(\beta_1, \beta_2)$ plane to obtain a more accurate value. We have chosen trajectories that cross the transition line nearly orthogonally. For point **D** we have moved along the line $\beta_1 = 2$. For point **E** we have chosen the line $\beta_1 + \frac{1}{2}\beta_2 = 0$. In the case of point **F** the simulations have been done at $\beta_1 = 0.5$, moving along a line $\beta_1 - \frac{1}{2}\beta_2 = \text{constant}$ .

The Finite Size Scaling of the specific heat fits to a behavior of type (19) only for very large lattices. A better fit can be obtained to a dependence of type

$$C \sim A + BL^{\alpha/\nu}, \tag{24}$$

where $A$ and $B$ are constants. To avoid a difficult three parameter fit to determine $\alpha/\nu$ one could carry out a linear fit of $\log C$ as a function of $\log L$. However this may be a very dangerous procedure, due to the presence of the constant term $A$, that would require very large values of $L$ in order to find a clear asymptotic behavior. In figure 2 this bilogarithmic plot shows the lack of asymptotic behavior in lattices as large as $L = 64$.

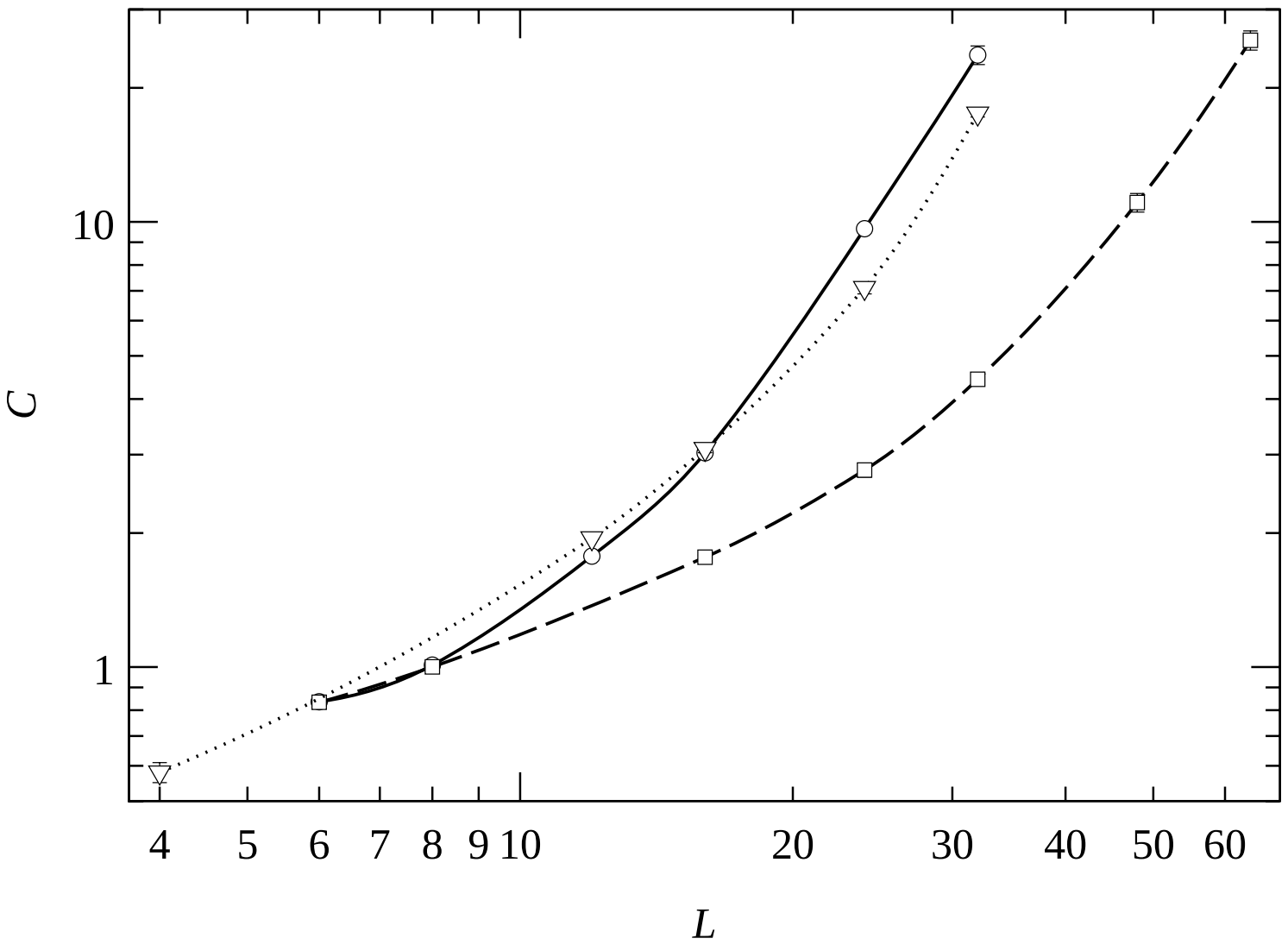

Figure 2: Specific heat as a function of the lattice size in a bilogarithmic scale for the points **D** (circles), **E** (squares) and **F** (triangles). The error bars are smaller than the symbol sizes.

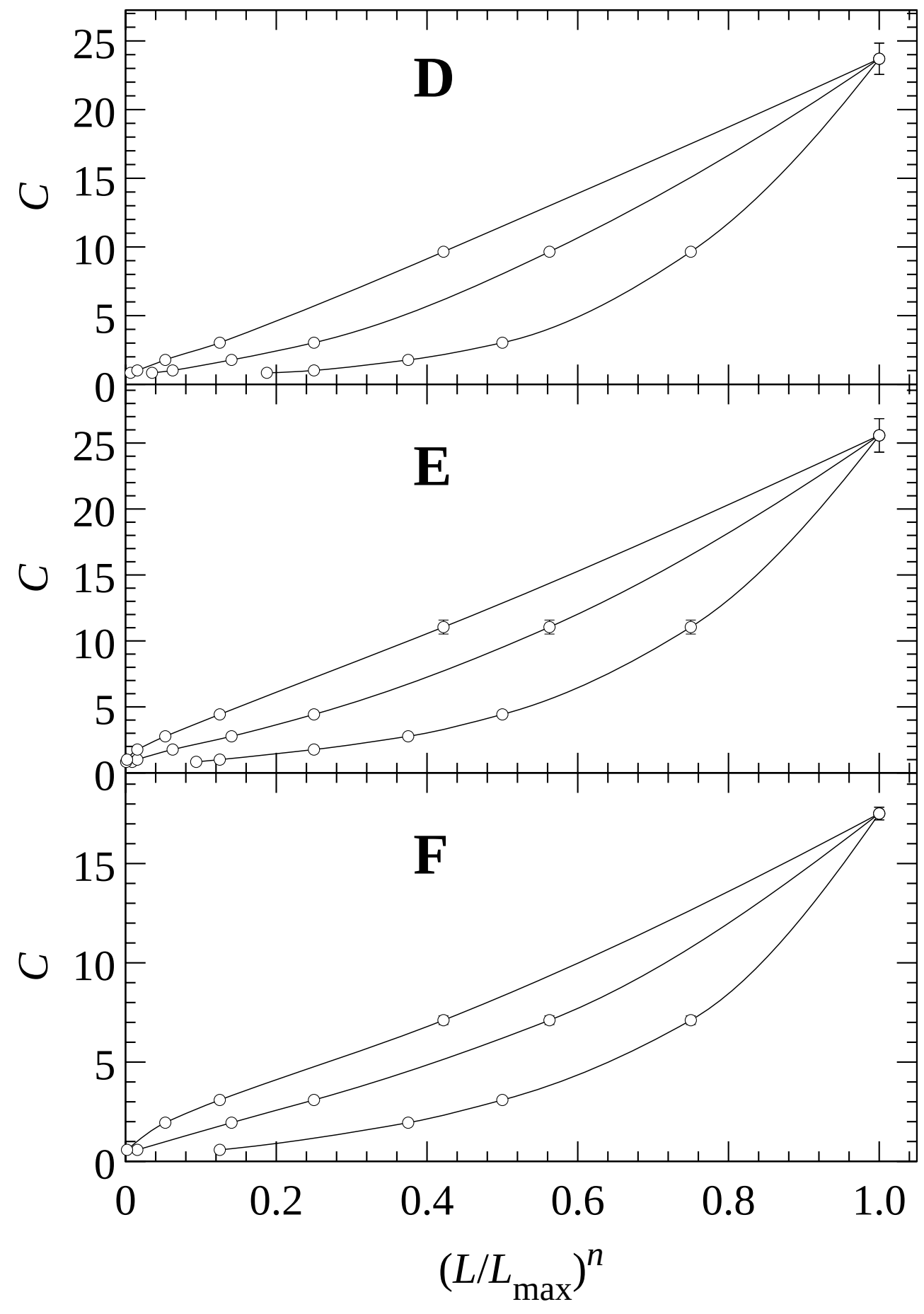

Figure 3: Specific heat vs. $(L/L_{\rm max})^n$ $n = 1, 2, 3$ in the points **D**, **E** and **F**. We see that the specific heat grows with the volume of the lattice.

In figure 3 we plot $C$ as a function of several powers of $L$. It is clear that for $n = 3$ (corresponding to $\alpha/\nu = 3$ ) the behavior is almost linear with a nonvanishing $L \to 0$ limit.

We should emphasize that the asymptotic (first order) behavior is reached in all cases but it is more difficult to see at the point labeled **E**.

Another interesting quantity is the magnetic susceptibility. In this case, usually it is not necessary to add a constant term to the power law (20) to obtain a good linear behavior for reasonable lattice sizes. The results are summarized in figure 4. We observe an absence of linear behavior even for the larger lattices. If the slope is computed with contiguous points we obtain values greater than 3 for the larger lattices. We conclude that this quantity is badly behaved because of the first order nature of the transition. In fact it is defined as the difference of two functions that become discontinuous in the $L \to \infty$ limit. Nevertheless, the behavior found excludes a second order transition.

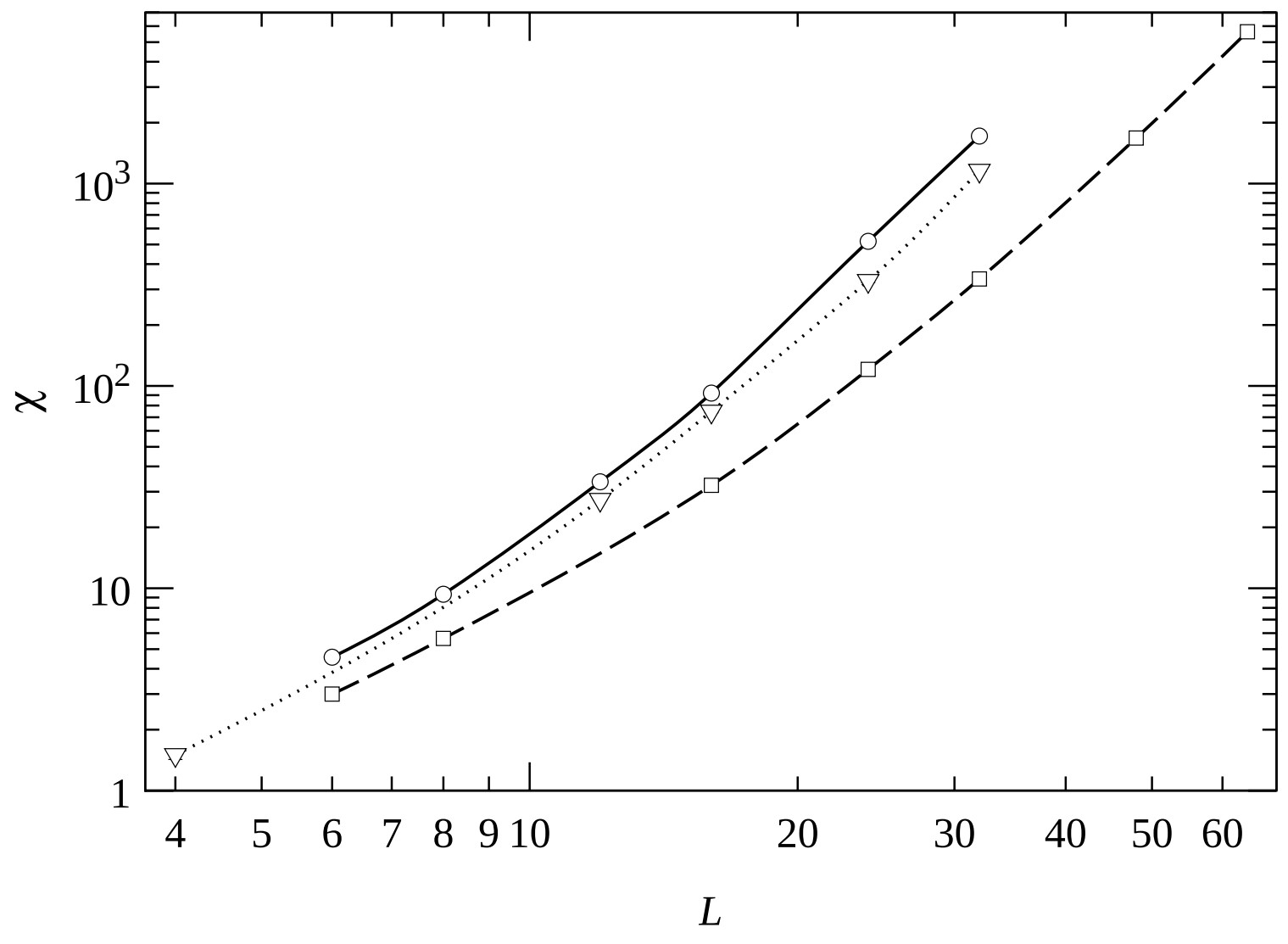

Figure 4: Susceptibility as a function of the lattice size in a bilogarithmic scale for the points **D** (circles), **E** (squares) and **F** (triangles).

Regarding the energy histograms, the results for the three points are shown in figure 5. The first order behavior (two peak structure) is again conclusive, with a stable inter peak distance for growing lattice size and a decreasing height in the intermediate region. It is interesting to point out that even when $L$ is too small to resolve both peaks the analysis of the specific heat or of the susceptibility gives strong indications of a first order character of the transition.

Finally we consider the correlation length defined in (16). In figure 6 (left) we plot $\xi$ as a function of $\beta_2$ for several lattice sizes in the point **D**. Only for $\beta_2 > \beta_2^c$ the plotted quantity makes sense as a correlation length. The value at the critical point has to be obtained computing this point by other means.

Notice that the lowest line, corresponding to the $L = 48$ lattice, does not jump at the transition. The reason for this is that the simulation was carried out in the disordered phase and then extrapolated, so it accounts only for the metastable disordered state.

The coincidence with the $L = 32$ plot in the unbroken phase is very reassuring, in the sense that we have an accurate measure of the correlation length, which seems fairly stable until the transition point.

Following the same procedure for points **E** and **F**, middle and right parts of figure 6, we quote

$$\begin{aligned} \xi_{\max}^{D} &\sim 7\ , \\ \xi_{\max}^{E} &\sim 12\ , \\ \xi_{\max}^{F} &\sim 7\ , \end{aligned} \tag{25}$$

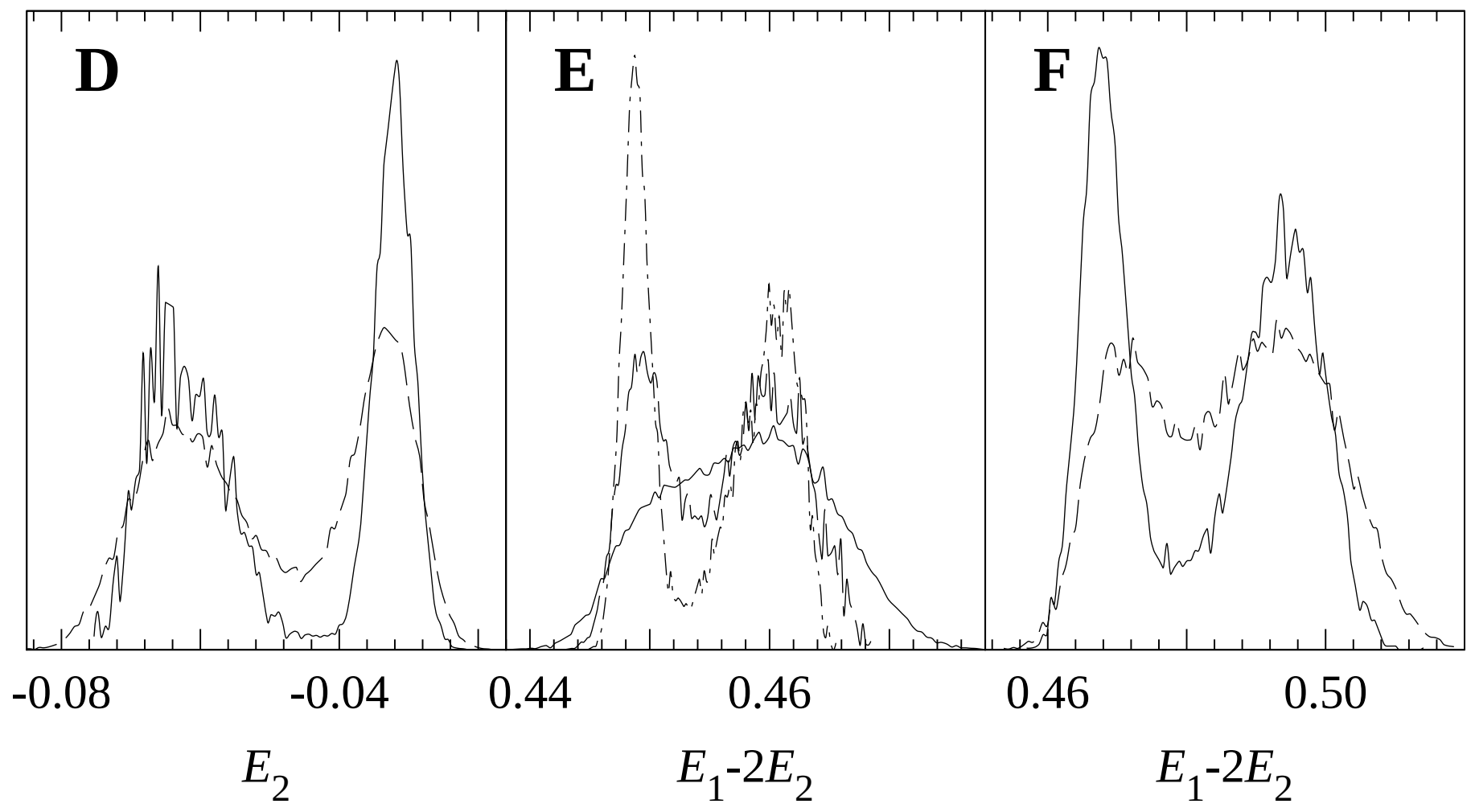

Figure 5: Energy histograms for $L = 24$ (dashes), $L = 32$ (solid), $L = 48$ (long dashes), and $L = 64$ (dot-dashes) at points **D**, **E** and **F**.

where the statistical errors are about a 10%. The values in (25) give an *a posteriori* explanation of the different levels of difficulty in reaching the asymptotic behavior in each case.

The shift of the critical point is too small for extracting the critical exponent $\nu$ with accuracy, but for the larger lattices it is compatible with an $L^{-3}$ behavior. Linear fitting the apparent critical points for the larger lattices, as a function of the inverse of the volume, we obtain the following infinite volume values for the selected transition points:

$$\begin{array}{rclrcl} \beta_1^{\mathbf{D}} &=& 2\ , & \beta_2^{\mathbf{D}} &=& -1.25111(13)\ , \\ \beta_1^{\mathbf{E}} &=& 0.85763(8)\ , & \beta_2^{\mathbf{E}} &=& \beta_1^{\mathbf{E}}/2\ , \\ \beta_1^{\mathbf{F}} &=& 0.5\ , & \beta_2^{\mathbf{F}} &=& -2.3899(12)\ . \end{array} \tag{26}$$

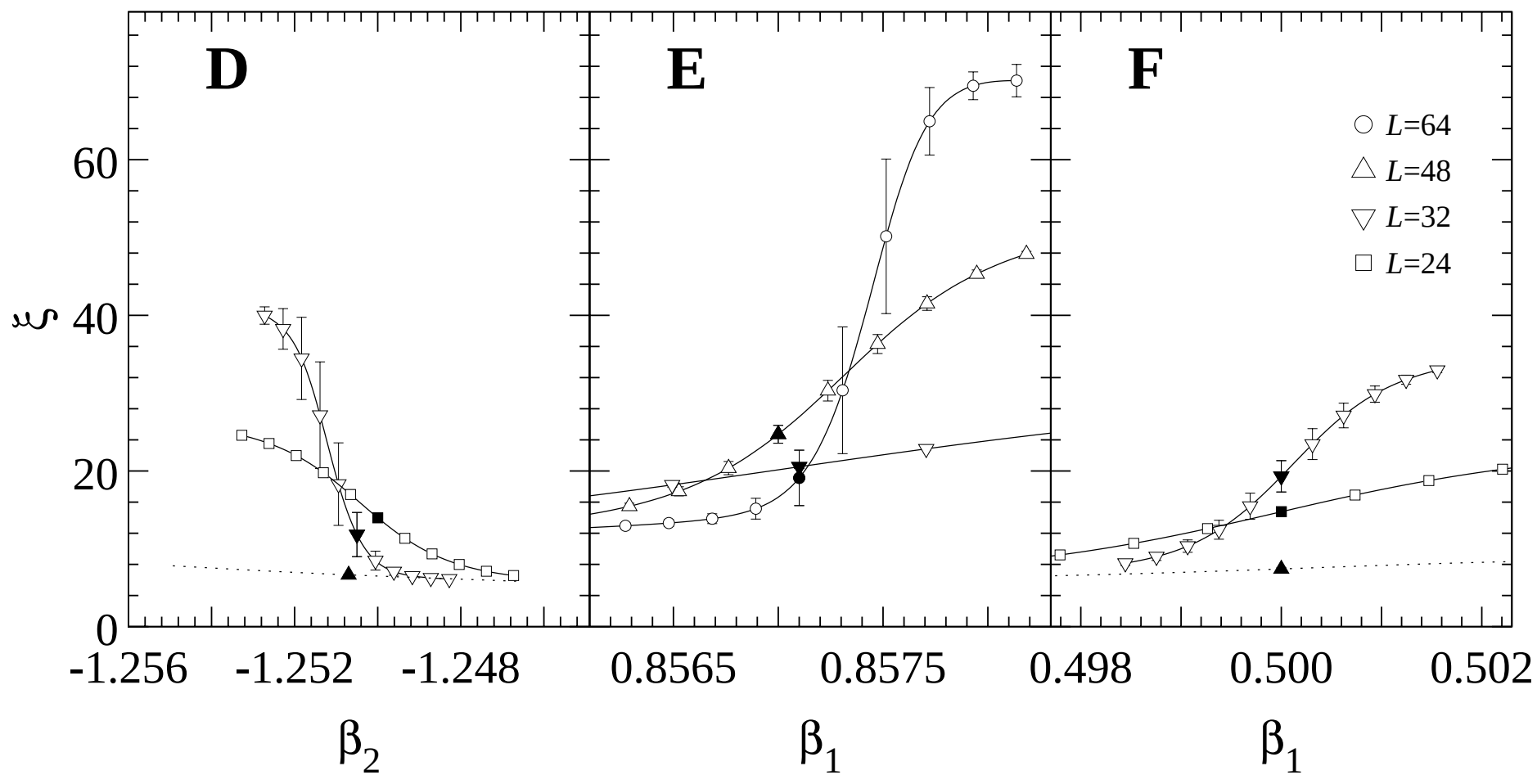

Figure 6: Correlation length as a function of one coupling at points **D**, **E**, and **F**. The dotted lines corresponds to disordered metastable states. Filled symbols have been plotted at the simulation point and white ones have been obtained with the spectral density method moving along lines $\beta_1 = 2$ for point **D** and $2\beta_1 + \beta_2 =$ constant for points **E** and **F**.

## 5.2 The FCC lattice

In this model, when $\beta$ is large and negative, the ground state becomes very complex. In addition to the global symmetries, the classical ($T = 0$) ground state is continuously degenerated with a $O(3)^L$ degeneracy group: the lattice will exfoliate in planes perpendicular to one of the lattice axis. We can arbitrarily rotate the privileged direction of every plane, without changing the energy (see figure 7). This is a very common feature of vector spin systems with competing couplings.

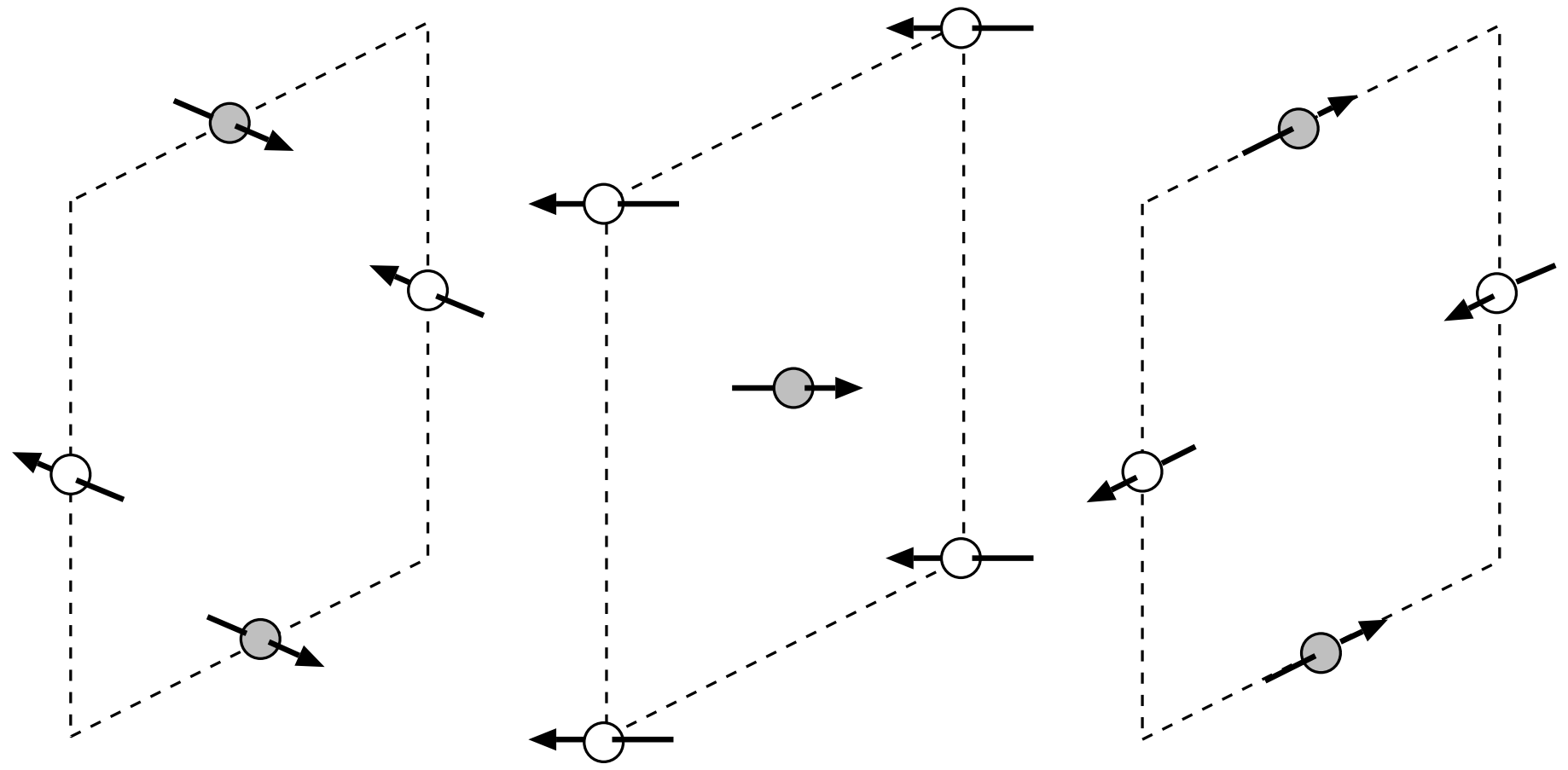

Figure 7: Classic ground state for the FCC lattice. In the figure, the distance between planes has been stretched out for clarity. A global rotation of the spins of any plane does not change the total energy.

However the $O(3)^L$ ground state invariance is not a symmetry of the Hamiltonian and then we expect the thermal fluctuations to select a discrete subset as the stable ground state. This is known as Villain's *order from disorder* [21]. A first order free energy calculation in the spin wave approximation [22] indicates that the *collinear* ground state is selected: that is one in which the $O(3)^L$ symmetry is broken and all spins in all planes are oriented in the same direction.

Let us see how this general tendency can be understood [22]: the local magnetic field acting on a spin in a given plane (consider, for instance, the black spin in the middle plane in figure 7) is the sum of the magnetic fields of its twelve nearest neighbors. In any ground state at zero temperature, the magnetic field will be null, but thermal fluctuations in each plane will produce a magnetic field perpendicular to its privileged direction. The spin in the middle plane will then orientate orthogonally to the mean magnetic field so that the spin fluctuations will be parallel to it, selecting a *collinear* ground state. In the absence of another interaction that unambiguously fixes the ground state (in ref [22] they consider also a second neighbor interaction), we still have a remaining $Z_2^L$ degeneracy, as, according to the previous heuristic argument, all the spins in a given plane may be inverted. We have carried out a free energy calculation in the spin wave approximation that shows that, up to first order, the degeneracy is unbroken. Specifically, let us consider any one of these $2^L$ ground states, $\{\boldsymbol{\sigma}_i^0\}$. Any configuration near the ground state can be written $\{\sqrt{1-\boldsymbol{\pi}^2}\,\boldsymbol{\sigma}_i^0 + \boldsymbol{\pi}_i\}$ with $\boldsymbol{\sigma}_i^0 \cdot \boldsymbol{\pi}_i = 0$. Regarding the partition function, it can be shown in perturbation theory that the lower order term that could depend on the specific collinear ground state selected, is $O(\boldsymbol{\pi}^8)$.

The numerical analysis of the low temperature regime by means of Monte Carlo simulation is to be interpreted with great care, because finite size effects may be very misleading.

We have performed a numerical simulation with an $L = 24$ lattice in the low temperature phase ($\beta = -5 < \beta_c$) that confirms the collinear prediction ($\mathrm{O}(3)^L$ breaking). However, every one of the $2^L$ ground states is very stable under Monte Carlo evolution with a local update algorithm in the sense that the system does not move from a neighborhood of the corresponding ground state. On the other hand, we have computed the mean energy starting from different configurations, and we have obtained the same values within errors. The errors are about $10^{-5}$ to be compared with the latent heat of the transition ($\sim 10^{-2}$).

Therefore, a proper simulation in the low temperature phase becomes very hard given the very long tunneling time. This is of no significance for real magnetic crystals like $\mathrm{UO}_2$, because any next nearest neighbors coupling, breaks this degeneracy.

There is a previous work [23], where the model has been studied, and evidence for the first order nature of the transition is presented. However, they do not describe the $Z_2^L$ degeneration, although they report different results for their order parameter if the starting configuration was random or ordered (their ordered configuration was just one special case of the $2^L$ ground states, which would have been natural with a second neighbors coupling).

The picture we get from this is a disordered phase until $\beta \sim -4.5$ where a first order transition takes place to a very different phase, roughly collinear, but with a somehow glassy behavior.

In figure 8 we show a bilogarithmic plot of the maximum of the specific heat as a function of the lattice size. The growing for the larger lattices is even larger that the asymptotic value expected for a first order transition. This phenomenon is understood as a transient increasing of the latent heat. In figure 9 we show a very clear double peak histogram for the $L = 32$ lattice. Comparing with the $L = 24$ case we see a wider histogram in addition to the effect of the decreasing of the probability in the central region. Although we cannot consider the behavior at $L = 32$ as asymptotic, the first order nature of the transition is well established.

Using the larger lattices we have computed the extrapolation to infinite volume of the transition point, under the hypothesis of first order behavior, obtaining

$$\beta_c^{\mathrm{FCC}} = -4.491(2). \tag{27}$$

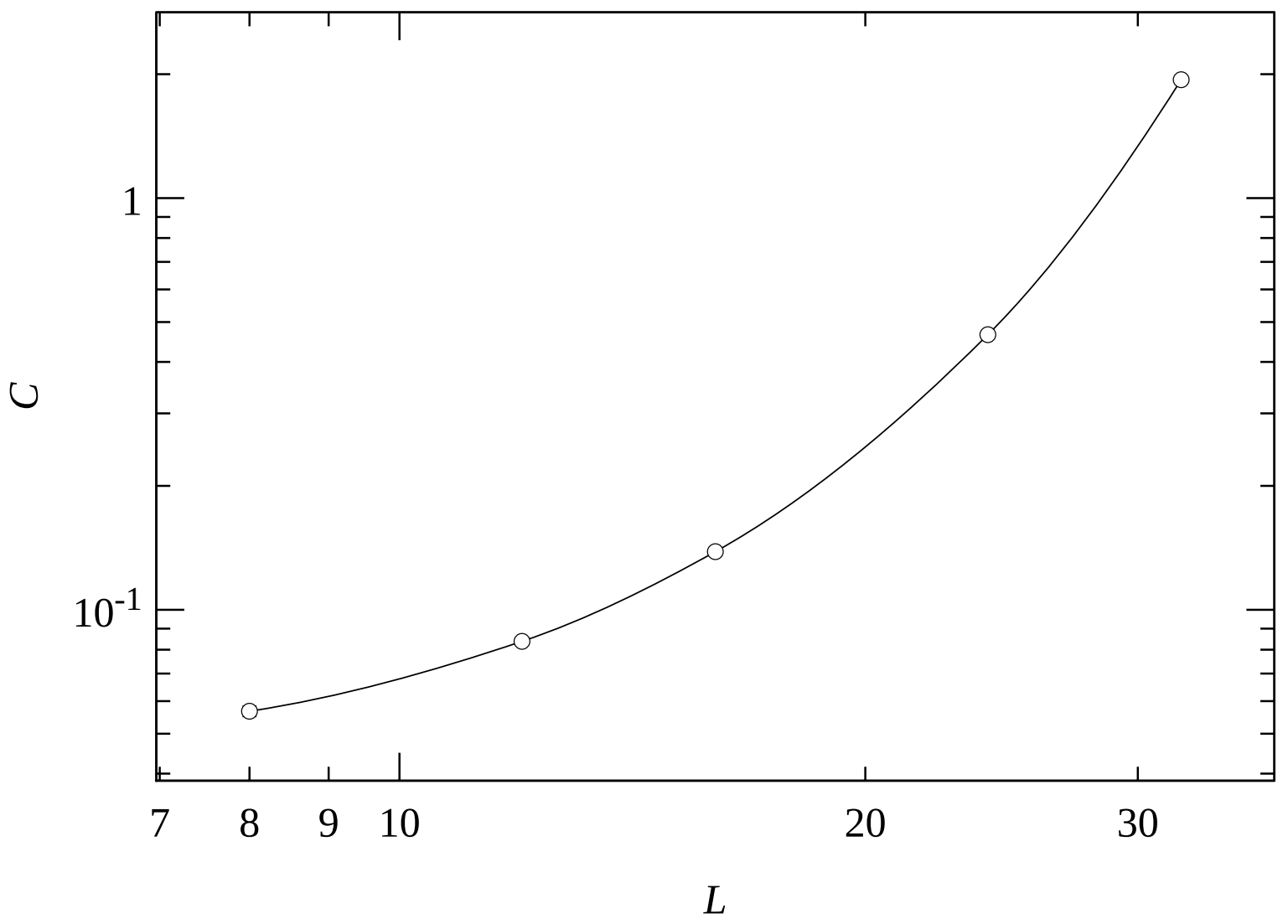

Figure 8: Specific heat as a function of the lattice size in a bilogarithmic scale for the FCC lattice. The slope of the segment joining points $L = 8$ and $L = 12$ is $0.96(7)$ and that corresponding to $L = 24$ and $L = 32$ is $4.96(8)$.

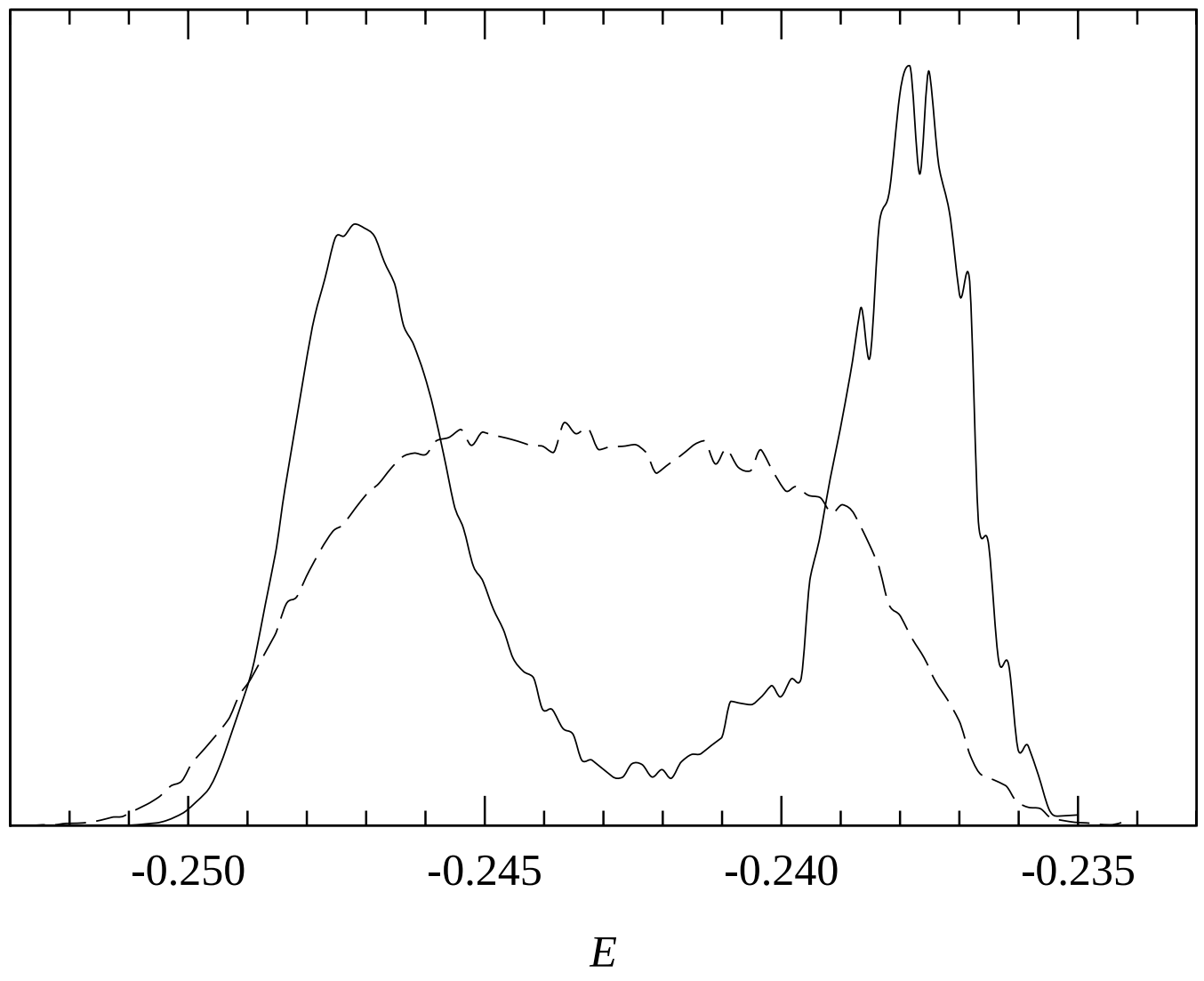

Figure 9: Energy histogram for $L = 24$ (dashed line) and $L = 32$ (solid line) at the antiferromagnetic transition in the FCC lattice.

## 5.3 The Fully Frustrated model

The Hamiltonian (see eq. (5)) of this model is invariant under the transformation (3) so we will consider only the $\beta \geq 0$ case.

At small $\beta$ value the model presents a disorder phase. At large $\beta$ the system becomes ordered, but with a fairly complicated structure. A phase transition is observed at

$$\beta_c^{\mathrm{FF}} = 2.26331(13) \tag{28}$$

Let us first discuss the properties of the ordered phase. We have found directly from the simulation that the modulus of the magnetization defined in (12) is large at $\beta > \beta_c$ and it goes to 1 when $\beta \to \infty$. This means that the system develops a period two structure, allowing us to characterize the vacuum just by studying the unit cell ($2^3$ spins) which we have done by means of analytical and numerical methods.

In the unit cell, with periodic boundary conditions, from the $8 \times 2$ parameters we can fix a direction on the internal space and an azimuthal angle, remaining 13 parameters. The ground state is highly degenerate: the set of minimal energy configurations is a two dimensional manifold.

Let us call $e_i \equiv \boldsymbol{\sigma}(\boldsymbol{r}_0) \cdot \boldsymbol{\sigma}(\boldsymbol{r}_i)$, where the site $\boldsymbol{r}_0 \equiv (0,0,0)$ and $\boldsymbol{r}_i$ is one of the sites $\boldsymbol{r}_1 \equiv (1,0,0)$, $\boldsymbol{r}_2 \equiv (0,1,0)$ $\boldsymbol{r}_3 \equiv (0,0,1)$. Due to the $Z_2$ local gauge transformation it can be assumed that the couplings between those sites are positive. It is easy to check that given any configuration with values $e_i = c_i$ there is another configuration with the same total energy for any permutation of $\{c_1, c_2, c_3\}$. We have generated different ground state configurations, whose total energy is $24 \times \frac{1}{\sqrt{3}}$, and we have checked that, in the corresponding three dimensional space $\{e_1, e_2, e_3\}$, they lie in a plane perpendicular to the vector $(1,1,1)$ filling an hexagon with vertices at $(1, \frac{2\sqrt{3}-3}{3}, \frac{\sqrt{3}}{3})$ and permutations.

In a $L > 2$ lattice at $\beta > \beta_c$ but finite, the equilibrium configurations concentrate in the six corners of the hexagon in a region whose size decreases for increasing lattice size. For example, at $\beta = 10$ in an $L \geq 16$ lattice the system is unable to flip between vertices in the local Monte Carlo evolution. We must point out that the selection of the vertex as the origin of the elementary cell is arbitrary. Selecting another origin the results are equivalent. We interpret the behavior of the ordered phase as a Villain's order from disorder mechanism.

It is rather natural, that such a complicated ordered phase, where entropy selects one of the continuously infinite ground states, will make the thermodynamic limit hard to reach. In fact, this is a weak first order transition as the finite size scaling of the specific heat shows (see the figures 10 and 11). Careful observation of figure 10 shows that for lattice sizes from $L = 8$ to $L = 24$ we find a fairly good fit to $\alpha/\nu = 1$. This is a typical signature of a weak first order transition [24]. Indeed, bigger lattices show a completely different finite size scaling behavior: for $L = 24$ to $L = 64$ we find $\alpha/\nu = 3$ as expected in a first order transition (see figure 10), although we only observe for $L = 64$ an incipient two peak structure in the energy histogram.

For the susceptibility, we find traces of the weak first order character of the transition, in the growing slope of the bilogarithmic plot of the maximum value of the susceptibility versus the lattice size (see figure 12). For the biggest lattices, it is apparent that $\gamma/\nu \sim 3$.

Finally, by observing the evolution of the order parameter as $L$ grows (figure 13), we conclude that in the infinite volume limit it becomes a discontinuous function of $\beta$.

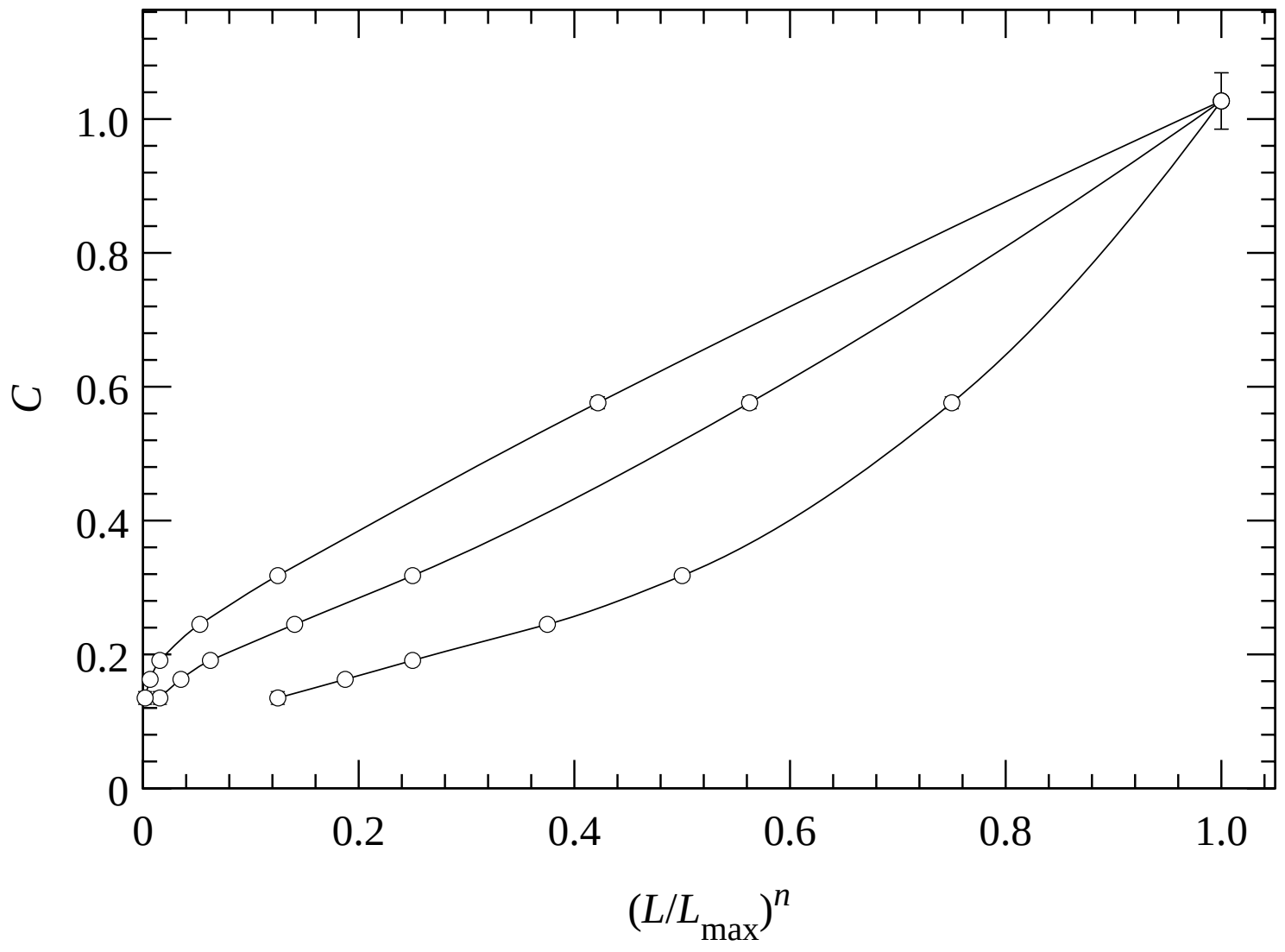

Figure 10: Specific heat vs. $(L/L_{\rm max})^n$, $n = 1, 2, 3$ in the FF model. We see that the specific heat grows with the volume of the lattice.

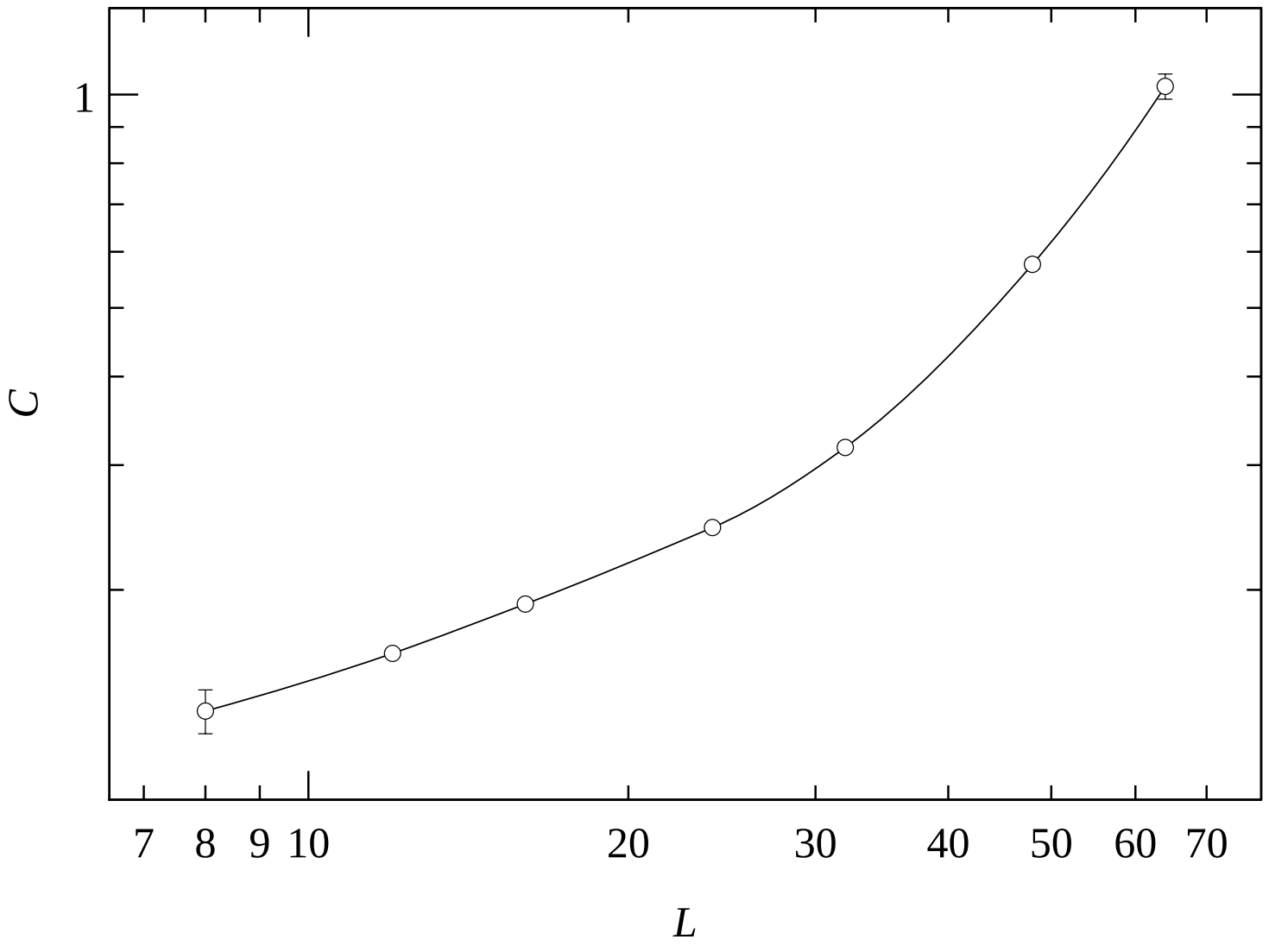

Figure 11: Specific heat as a function of the lattice size in a bilogarithmic scale for the FF lattice. The slopes computed from adjacent points ranges from 0.46(18) to 2.01(15).

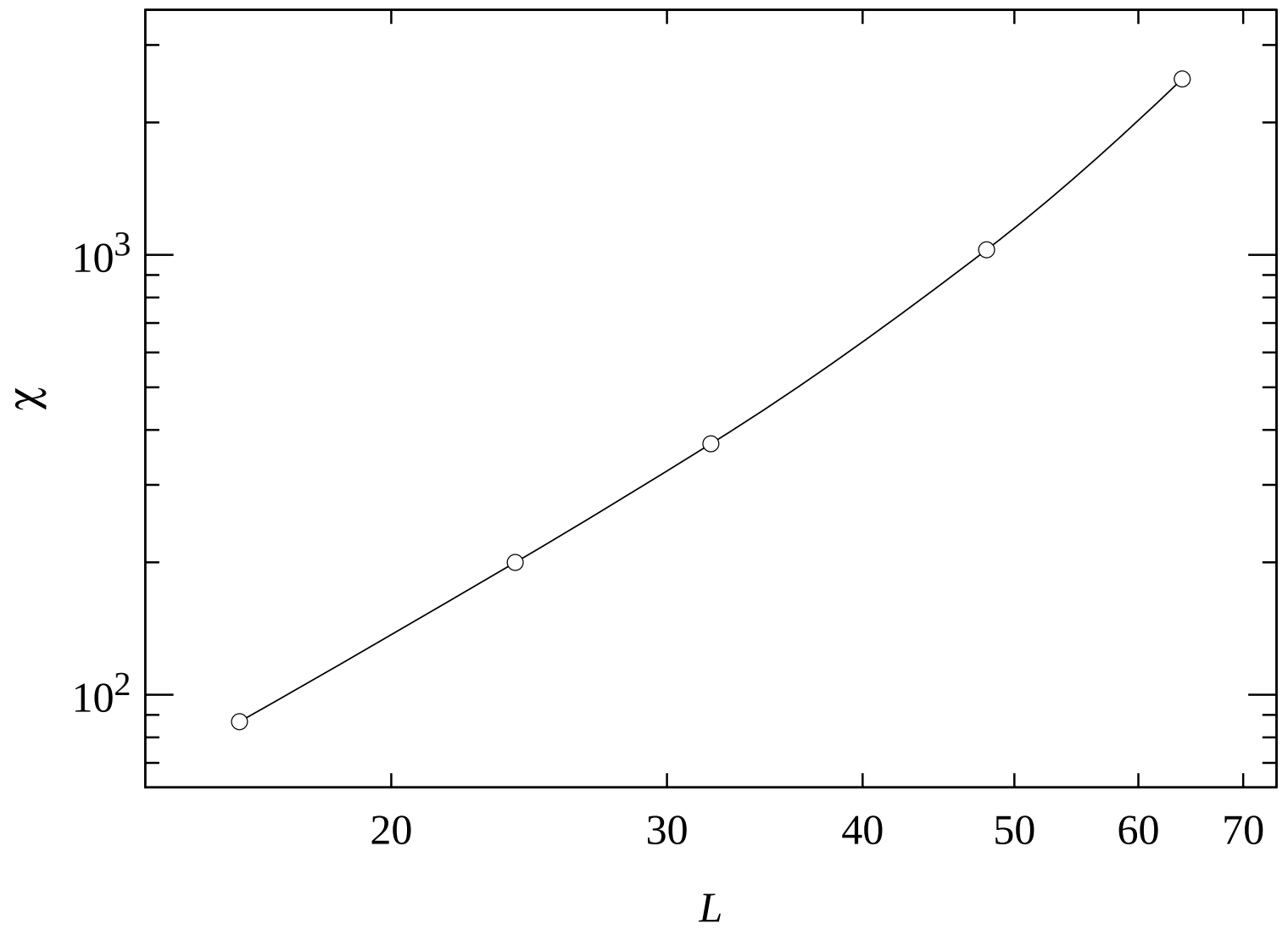

Figure 12: Susceptibility as a function of the lattice size in a bilogarithmic scale for the FF lattice. The slope from the two smallest lattices is 2.06(5) and from the two largest 3.11(9).

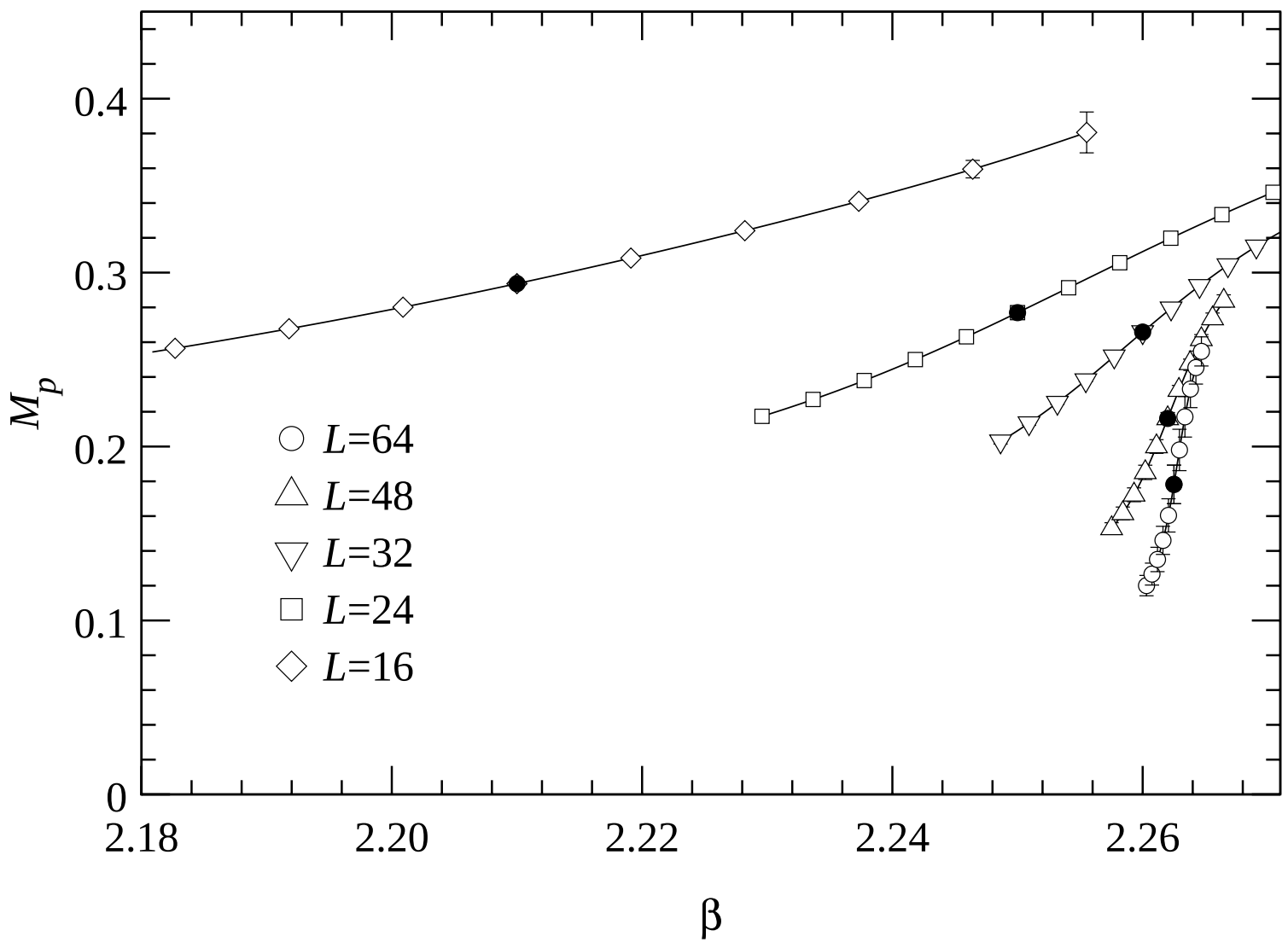

Figure 13: Period two magnetization vs. $\beta$ for different lattice sizes. The discontinuous character of the transition is striking in the $L \to \infty$ limit.

# 6 Conclusions

We have explored with Monte Carlo simulations three models with internal O(3) symmetry that develop frustrated antiferromagnetic vacua: a model with ferromagnetic nearest neighbors interaction and antiferromagnetic second nearest neighbors one; an antiferromagnetic model in a FCC lattice; and a Fully Frustrated model. We found a very rich variety of vacuum structures. In particular we have observed different cases of Villain's order from disorder mechanism.

By studying the Finite Size Scaling behavior of the specific heat, susceptibility and correlation length we conclude that all the antiferromagnetic transitions found are of first order. This conclusion is also confirmed with the analysis of the energy distribution of the equilibrium configurations. In some cases lattices as large as $48^3$ or $64^3$ have been needed to observe a double peak structure of the energy histogram.

On the other hand, we have checked that all the ferromagnetic line in the two parameter model as well as the FCC limit belong to the same universality class. This seems to be the only universality class for three dimensional O(3) models.

### Acknowledgements

We thank to Gabriel Álvarez Galindo, Alan Sokal, and Arjan Van der Sijs for useful discussions. Partially supported by CICyT AEN93-0604, AEN94-0218 and AEN93-0776. H. G. Ballesteros and V. Martín-Mayor are MEC fellows.